\documentclass[twocolumn,prd,showpacs,nofootinbib,preprintnumbers,
               amsmath,amssymb]{revtex4}
\usepackage{graphicx}

\newcommand{\Real}{\mathop\mathrm{Re}\nolimits}
\newcommand{\Imag}{\mathop\mathrm{Im}\nolimits}
\newcommand{\half}{\frac{1}{2}}
\newcommand{\sm}[1]{{\scriptscriptstyle \rm #1}} 
\def\la{\langle }
\def\ra{\rangle }  
\renewcommand{\l}{\left(}
\renewcommand{\r}{\right)}

\begin{document}

\preprint{INR/TH-2003-4}

\title{Transverse muon polarization in $K^+\to\mu^+\nu_\mu\gamma$: 
       scanning over the Dalitz plot}
\author{F.~L.~Bezrukov}
  \email{fedor@ms2.inr.ac.ru}
\author{D.~S.~Gorbunov}
  \email{gorby@ms2.inr.ac.ru}
\author{Yu.~G.~Kudenko}
  \email{kudenko@wocup.inr.troitsk.ru}
\affiliation{
    Institute for Nuclear Research of the Russian Academy of Sciences,\\
    60th October Anniversary prospect 7a, Moscow 117312, Russia
}
\date{April 15, 2003}

\begin{abstract}
We study the potential of the
measurement of the transverse muon polarization $P_T$ in the
$K\to\mu\nu_\mu\gamma$ decay with the  sensitivity of 
$\delta P_T\sim 10^{-4}$. It is shown that the forthcoming
experiment can measure the contribution of the electromagnetic final
state interactions to $P_T$ that gives a possibility to
unambiguously determine the signs of the sum of the kaon form factors
$F_V$ and $F_A$ even without fixing their difference.  We also
estimate the sensitivity of this experiment to the new physics, which
could give rise to $T$-violation: multi-Higgs doublet models,
supersymmetric extensions of the Standard Model, left-right symmetric
model and leptoquark models.
\end{abstract}     


\maketitle

\section{Introduction}

In spite of remarkable progress in  study of $CP$-violation phenomena
in both K and B sectors it remains an interesting issue.  The Standard
Model~(SM) successfully describes existing experimental data by a single 
phase of the CKM matrix, although it is hard to believe that this
phase is the only source of $CP$-violation.  For example, the baryon
asymmetry of the Universe can not be explained by the CKM phase only
and at least one additional source of the $CP$-violation is required.

A good place to look for new $CP$-violating phases is a measurement of
processes where the SM $CP$-violation is vanishing or very
suppressed while additional or alternative sources of $CP$-violation can
produce a sizable effect.  Such interesting observables are the
electric dipole moment of neutron which is extremely small in the SM,
and the transverse lepton polarization  in three-body decays
of kaons and B-mesons~\cite{Lee:iz,general,Gabrielli:1992dp}.

In this paper, we study the $T$-odd muon polarization in the decay
$K^+\to\mu^+\nu_\mu\gamma$~($K_{\mu2\gamma}$). Namely, we investigate the
possibility to measure the vector and axial-vector kaon form factors,
$F_V$ and $F_A$, from the $T$-odd muon polarization emerging due to the
electromagnetic final state interactions (FSI). Also we analyze potential
contributions to $P_T$ from various  extensions of the SM. In this study 
we focus on the 
Dalitz plot region where $K_{\mu2\gamma}$ events have a large angle between 
photon and muon momenta. 
This is the region, where $T$-odd muon polarization exhibits the best
sensitivity to $F_V$ and $F_A$. Moreover, we found that in this region
the ratio of the possible contribution to $P_T$ from new physics and
FSI contribution to $P_T$ becomes the largest. Hence, the analysis of
the $K_{\mu2\gamma}$ events at large $\theta$ will provide the best
accuracy in measurement of the relevant
parameters of the new physics or the strongest constraints on them. 

It is found that $P_T$ dependence on the kaon form factors gives a
possibility to unambiguously determine in $K_{\mu2\gamma}$ 
decay the signs of the sum of the kaon form factors 
without fixing their difference. 
Combined with 
measurement of the normal muon polarization, which is very sensitive
to $F_V$ and $F_A$~\cite{Bezrukov:2003ka}, this allows 
the values of the kaon form factors to be unambiguously
extracted with 1\% accuracy from this experiment. 
Investigating the prospects of new experiments in searching for new
physics we show that, generally, they are limited mostly by the
uncertainty in FSI predictions rather than by the anticipated
statistical error. 

The outline of the paper is as follows.  In Section~\ref{SecII} we introduce
the parameters describing muon transverse polarization $P_T$ in the
$K_{\mu2\gamma}$ decay, recall the relevant formulae and 
estimate the expected sensitivity to $P_T$ in forthcoming experiments.
In Sec.~\ref{SecIII}, an approach to determine the signs of the sum of
the kaon form factors from the FSI polarization is discussed
and the accuracy of this method is estimated.  In Sec.~\ref{SecIV} 
we study the discovery potential of the $P_T$ measurements in search for new
physics. Specifically, we estimate the upper bounds on the $P_T$ in 
multi-Higgs doublet models, supersymmetric
  extensions of the Standard Model, left-right symmetric model and 
leptoquark models. Sec.~\ref{SecV} contains conclusions and
final remarks.

\section{Transverse muon polarization in $K_{\mu2\gamma}$ decay}
\label{SecII}

\subsection{General description}

Introducing three unit vectors
\begin{equation*}
  \vec{e}_L= \frac{\vec{p}_\mu }{ |\vec{p}_\mu|} \,,\;
  \vec{e}_N= \frac{\vec{p}_\mu\times (\vec{q}\times \vec{p}_\mu)
    }{ |\vec{p}_\mu\times (\vec{q}\times \vec{p}_\mu)|} \,,\;
  \vec{e}_T = \frac{\vec{q}\times \vec{p}_\mu}{
    |\vec{q}\times \vec{p}_\mu|} \,,
\end{equation*}
with $p_\mu$ and $q$ being the four-momenta of $\mu^+$ and $\gamma$,
respectively, one can define longitudinal ($P_L$), normal ($P_N$) and
transverse ($P_T$) components of the muon polarization as the
corresponding contributions to the squared matrix element of
the $K_{\mu2\gamma}$ decay,
\begin{equation*}
  |M|^2=\rho_{0}[1+(P_L {\vec e}_L+P_N {\vec e}_N+P_T {\vec
  e}_T)\cdot \vec{\xi}\,]\;,
\end{equation*}
with ${\vec \xi}$ being a unit vector along muon spin and $\rho_{0}$
is 
\begin{multline*}
  \rho_0(x,y) = \frac{1}{2}e^2 G^2_F |V_{us}|^2 
    (1-\lambda) \times\\
  \left\{ 
  f_\mathrm{IB}(x,y)+f_\mathrm{SD}(x,y)+f_\mathrm{IBSD}(x,y) \right\} \;,
\end{multline*}
where the internal bremsstrahlung ($\mathrm{IB}$), structure dependent 
($\mathrm{SD}$) and interference contributions ($\mathrm{IBSD}$) are given as 
follows~\cite{Gabrielli:1992dp,Bijnens:1992en,Chen:1997gf} 
\begin{align}
  \label{fIB}
  f_\mathrm{IB}&=\frac{4m^2_\mu|f_K|^2}{ \lambda x^2}
    \left[x^2+2(1-r_\mu)\left(1-x-\frac{r_\mu}{\lambda}\right)\right]\!,
  \\
  \label{fSD}
  f_\mathrm{SD}&= m^4_K x^2\bigg[|F_V+F_A|^2\frac{\lambda^2 }{ 1-\lambda}
      \left(1-x-\frac{r_\mu}{ \lambda}\right) \nonumber \\
    &\qquad+|F_V-F_A|^2(y-\lambda)\bigg]\;,
  \\
  \label{fIBSD}
  f_\mathrm{IBSD}&= - 4m_K m^2_\mu\bigg[\Real[f_K(F_V+F_A)^*]
      \left(1-x-\frac{r_\mu}{\lambda}\right) \nonumber\\
    &\qquad -\Real[f_K(F_V-F_A)^*]\frac{1-y+\lambda}{\lambda}\bigg] \;.
\end{align}
Here we used the standard notations $\lambda=(x+y-1-r_\mu)/x$,
$r_\mu=m^2_\mu/m^2_K$, and $x=2E_{\gamma}/m_K$, $y=2E_\mu/m_K$ with
$E_\gamma$, $E_\mu$ being the photon and muon energies in the kaon
rest frame, respectively; $G_F$ is the Fermi constant, $V_{us}$ is the
corresponding element of the Cabibbo--Kobayashi--Maskawa (CKM) matrix
and $f_K=159.8$~MeV is the kaon decay constant.  In terms of these variables 
the differential decay width reads
\begin{equation*}
  d\Gamma(\vec{\xi}\,) =
    \frac{m_K}{32(2\pi)^3}|M(x,y,\vec{\xi}\,)|^2 dx\,dy \;.
\end{equation*}
The transverse muon polarization $P_T$  is determined using 
the partial decay width
\begin{equation}\label{PT}
  P_T=\frac{d\Gamma(\vec{e}_T)-d\Gamma(-\vec{e}_T) }
           {d\Gamma(\vec{e}_T)+d\Gamma(-\vec{e}_T)}
     \equiv\frac{\rho_T}{\rho_0} \;.
\end{equation}
This is a $T$-odd observable (both $P_L$ and $P_N$ are $T$-even), hence in
the $T$-invariant theory its value equals zero at tree level.
Moreover, $P_T$ does not have tree-level contributions from the
$CP$-violating phase in the CKM matrix.  These two features make $P_T$
a very promising observable for new $CP$-violating physics searches.
At the same time, as will be seen in the Sec.~\ref{SecIII}, analysis
of loop contributions from the Standard Model (FSI) is also of a
special physical interest.

\subsection{Experimental sensitivity to $P_T$}

For the following analysis let us estimate the level of precision
which could be achieved experimentally.

The running E246 experiment at KEK \cite{Abe:1999nc} dedicated for a
measurement of $P_T$ in the decay $K^+\to \pi^0\mu^+\nu$ has only a
limited sensitivity of about $10^{-2}$ to $P_T$ in $K_{\mu2\gamma}$
\cite{Kudenko:2000sc}.  There is a proposal for a new experiment in
which a statistical sensitivity $P_T\leq 10^{-4}$ in $K_{\mu2\gamma}$
can be reached~\cite{Kudenko:yk}.  The main features of this
experiment include a high resolution measurement of neutral particles
from $K_{\mu3}$ and $K_{\mu2\gamma}$ decays, usage of an active muon
polarimeter which provides information about stopped muons (stopping
point, momentum), positron direction, and also detects photons, and a
highly efficient photon veto system covering nearly $4\pi$ solid
angle.  This approach allows to accumulate $K_{\mu2\gamma}$ events for
all $\theta$ angles between photon and muon momenta due to efficient
photon veto detector. This system eliminates $K_{\pi2}$ decays which
are the main background source at large $\theta$.

Using parameters of the proposed detector~\cite{Kudenko:yk} and the
kaon beam intensity of about $10^7\;K^+$ per second expected at the
JHF~\cite{jhf} one can estimate the sensitivity to $P_T$ which could be
achieved in this  experiment.  With the analyzing power of the detector
$\sim 0.3$ and the kinematical attenuation factor $\sim 0.8$, the
expected sensitivity to $P_T$ in some region ${\cal R}$ of the Dalitz
plot can be expressed as
\begin{equation}\label{eq:pt}
  \delta P_T({\cal R}) \simeq \frac{1}
    {0.3\cdot0.8\sqrt{N_{K_{\mu2\gamma}}({\cal R})}}\;,
\end{equation}
where $N_{K_{\mu2\gamma}}({\cal R})$ is the number of
$K_{\mu2\gamma}$ events in the region ${\cal R}$. 
In this experiment, the most effective suppression of background
events are anticipated in the kinematic region with 
$E_{\gamma}>20$~MeV, $E_{\mu}>200$~MeV. There the number 
of $K_{\mu2\gamma}$ events accumulated for one year
running period is estimated to be $3\times 10^{10}$. 
In what follows we adopt these cuts as well as the number of events. 

\begin{figure}
  \begin{center}
    \includegraphics[width=\columnwidth]{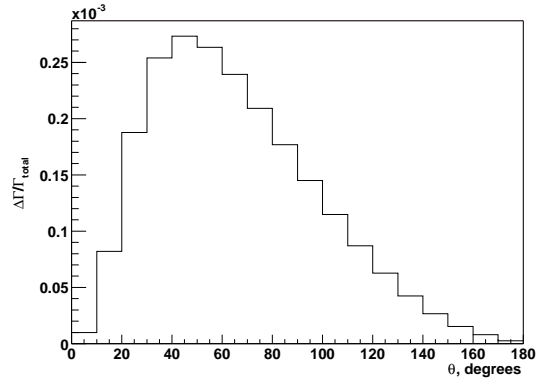}
  \end{center}
  \caption{Branching ratio for $K_{\mu2\gamma}$ decay for $10^\circ$
    wide bins in $\theta$; $E_\gamma>20$~MeV, $E_\mu>200$~MeV.  
    Dependence of this branching ratio on the values of the 
    $F_V$, $F_A$ form factors is quite weak.}
  \label{fig:2}
\end{figure}

The branching ratio of this decay  for $E_\gamma>20$~MeV and 
$E_\mu>200$~MeV integrated over $10^\circ$ wide bins
in $\theta$ is presented in Fig.~\ref{fig:2}.  
From this plot one can
estimate the sensitivity to $P_T$ in various $\theta$ regions.

\section{FSI and pinning down the kaon form factors}
\label{SecIII}

\subsection{Predictions and experimental data}

Due to lack of understanding of the QCD low-energy structure, there
is no definite prediction for the values of the $F_V$ and $F_A$ form
factors: the calculation of them is a model dependent procedure.  So,
the measurement of these form factors would give a possibility to make
a choice between various candidates for the correct description of the QCD
low-energy limit.

The distribution of the $K_{\mu2\gamma}$ decay width over Dalitz plot
allows  only the absolute values of the sum and difference
of the kaon form factors to be measured, since the term $f_\mathrm{IBSD}$ (see
Eq.~\eqref{fIBSD}) is small.  The terms with linear and quadratic
dependence on $F_V$ and $F_A$ give the comparable contribution in some
regions of the Dalitz plot that could, in principle, make it possible
to measure the signs as well as the magnitudes of the form factors.
Unfortunately, in the region where the linear terms grow, the dominant
contribution to $K_{\mu2\gamma}$ (internal bremsstrahlung \eqref{fIB},
which depends only on $f_K$) also increases, that significantly
reduces the sensitivity of Dalitz plot measurements
to the kaon form factors.  In practice, the situation is even worse,
since  only the absolute value of the sum of the
kaon form factors has been measured with good accuracy 
in these measurements, while their
difference still has only lower and upper
bounds~\cite{Adler:2000vk,Hagiwara:pw}:
\begin{gather}
  |F_V+F_A|=0.165\pm0.013\;,
  \label{bound-on-sum-from-Kl2gamma}\\
  -0.24<F_A-F_V<0.04 \;.
  \label{bound-on-dif-from-Kl2gamma}
\end{gather}

Recently, both vector and axial-vector form factors have been measured
in $K^+\to\mu^+\nu_{\mu}e^+e^-$ and $K^+\to e^+\nu_{e}e^+e^-$
decays~\cite{Poblaguev:2002ug}.  These decays are generalizations of
$K_{l2\gamma}$ for the case of a virtual photon in the final state, so
the kaon form factors $F_V$ and $F_A$ are believed to be the same in
all these processes.  The combined fit for both four-body decay
experiments results in
\begin{equation}\label{measured-values}
  F_V=-0.112\pm0.018 \;,\qquad F_A=-0.035\pm0.020 \;.
\end{equation}
These values are in a good agreement with ${\cal O}\l p^4\r$
predictions~\cite{CFT,Bijnens:1992en} of the chiral perturbation theory (ChPT)
\begin{equation}
  F_V=-0.096 \;,\qquad F_A=-0.041\pm0.006 \;.
  \label{CFT-predictions}
\end{equation}

\subsection{Dependence on $Q^2$}
\label{sec:Q2}

It is worth to note that actually  $F_V$ and $F_A$ are not
constants, but some functions of the momentum of the lepton pair,
$Q^2\equiv(p_K-q)^2$, with $p_K$ being kaon four-momentum.  
In ChPT $Q^2$-dependence emerges due to higher
order corrections, which have not been calculated yet.  Their
magnitude can be  estimated as
\begin{equation*}
  \frac{\Delta F_{V,A}}{F_{V,A}}\sim \frac{Q^2}{m_{V,A}^2}
  =(1-x)\frac{m_K^2}{m_{V,A}^2}\;,
\end{equation*}
where $m_{V}$ and $m_{A}$ are masses of the first strange hadronic
vector ($K^{\ast}$) and axial-vector ($K_1$) resonances, respectively.
This estimate implies corrections as large as 25\%, hence the
experimental data on $K_{l2\gamma}$ should be fitted with at least two
additional parameters, that generally decreases the chances to
determine $F_V$ and $F_A$ in $K_{l2\gamma}$ experiments.

In $K^+\to l^+\nu_{l}e^+e^-$ both vector and axial-vector form factors
get additional dependence on the non-zero $q^2$ that also increases
the bias in the corresponding fitting procedure.  In
Ref.~\cite{Poblaguev:2002ug} experimental data was fitted assuming
constant form factors and form factors depending on $Q^2$ and $q^2$
\begin{equation}\label{fQq}
  F_{V,A}(Q^2,q^2) = \frac{F_{V,A}}{(1-q^2/m_{\rho}^2)(1-Q^2/m_{V,A}^2)}
  \;,
\end{equation}
with $m_\rho$ being mass of $\rho$ meson.  This dependence appears in
the approximation of dominance of the contribution of lowest lying
resonances~\cite{CFT}.  
The experimental data favors the 
dependence~(\ref{fQq}) over the constant $F_{V,A}$, though the
accuracy achieved in this experiment did not allow definite
confirmation or rejection of $Q^2$-, $q^2$-dependence. 

The determination of the $Q^2$-dependence of the form factors may be
done in future experiments.  In particular, analysis of the $P_N$ muon
polarization in $K_{\mu2\gamma}$ decays can improve knowledge about
the $F_{V,A}$ behavior for sure~\cite{Bezrukov:2003ka}. 
However, at present, the unknown $Q^2$-dependence imports some
uncertainty into predictions for muon asymmetry to be measured in
future experiments. This concerns FSI contribution as well as possible
contribution from new physics, since both are functions of the kaon
form factors. 

To avoid all these difficulties one can try to find a physical
observable which strongly depends on $F_V$ and $F_A$ and can be
measured with an accuracy sufficient to distinguish the signs of kaon
form factors in the $K_{\mu2\gamma}$ decay. This implies that:
\emph{(i)} the observable we are interested in should exhibit linear
dependence on $F_V$ and $F_A$ in some region of the Dalitz plot;
\emph{(ii)} in this region the differential partial width should be
sufficiently large; \emph{(iii)} 
the $Q^2$-corrections in this region
should be small enough to unambiguously determine the sign of the kaon 
form factors.

Below we consider the transverse muon polarization calculated in the
framework of the Standard Model as a physical observable, which pins
down the signs of kaon form factors in $K_{\mu2\gamma}$ decay.

\subsection{Final state interactions}

\begin{figure*}
  \begin{center}
    \includegraphics[width=\textwidth]{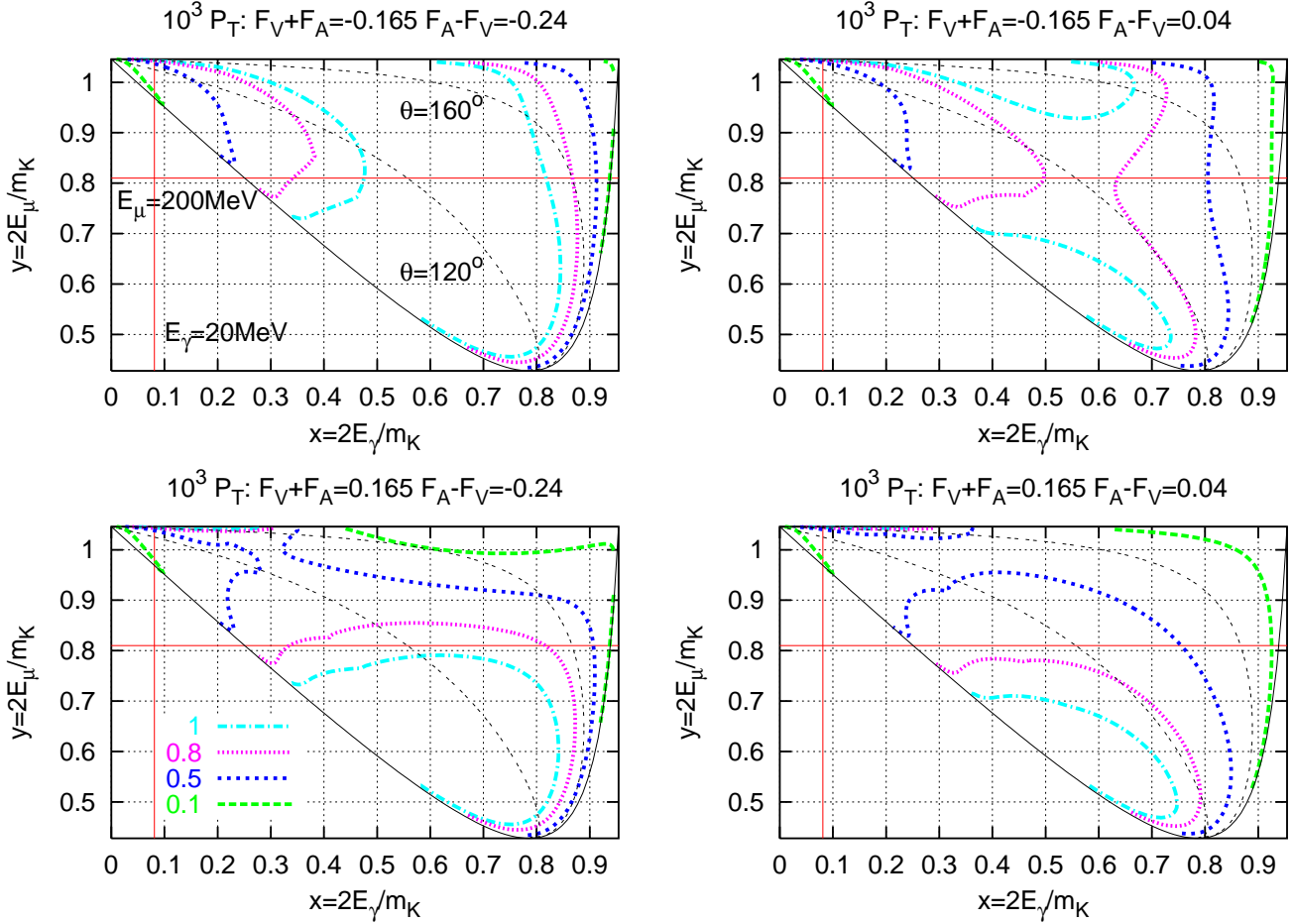}
  \end{center}
  \caption{$P_T$ distribution for different $F_V+F_A$ and $F_A-F_V$ values.
    The imposed cuts $E_\gamma>20 \mathrm{MeV}$,
    $E_\mu>200 \mathrm{MeV}$ and the region
    $120^\circ<\theta<160^\circ$ are also shown.}
  \label{fig:Pt_Dalitz}
\end{figure*}

In the framework of the Standard Model, the contribution to $P_T$
emerges only at loop levels due to  the FSI.  This 
contribution has been recently calculated~\cite{Braguta:2002gz} at
one-loop level. The expression for $\rho^{SM}_T(x,y)$ can be read out from
Ref.~\cite{Braguta:2002gz} after changing the definitions of the form
factors: $F^{\mbox{\scriptsize\cite{Braguta:2002gz}}}_V\to-m_KF_V$ and
$F^{\mbox{\scriptsize\cite{Braguta:2002gz}}}_A\to m_KF_A$.     
The asymmetry is positive
at any relevant $(x,y)$ and its absolute value ranges from zero to
$1.5\times10^{-3}$.  The Dalitz plot distribution of $P_T$ for several
values of the form factors satisfying
\eqref{bound-on-sum-from-Kl2gamma}, \eqref{bound-on-dif-from-Kl2gamma}
is presented in Fig.~\ref{fig:Pt_Dalitz}.

\subsection{Experimental prospects for determination of the form factors}

The transverse muon polarization emerging from the FSI exhibits 
the behavior we are looking for  
to pin down the signs of the kaon form
factors.  Indeed, the muon transverse polarization is sensitive to the
signs of $F_V$ and $F_A$, especially at large angles $\theta$ between
photon and muon momenta.

\begin{figure}
  \begin{center}
    \includegraphics[width=\columnwidth]{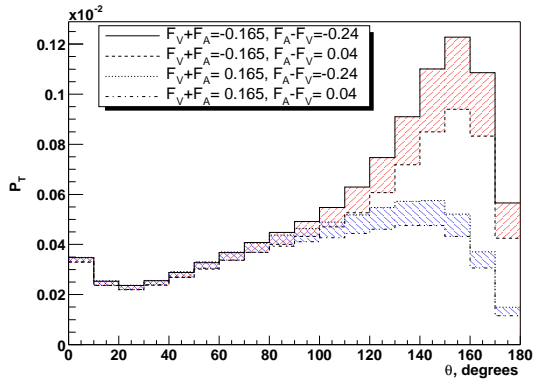}
  \end{center}
  \caption{$P_T$ (FSI) asymmetry for $10^\circ$ wide bins in $\theta$. 
    Upper region corresponds to negative $F_V+F_A$ and lower one --- to
    positive.}
  \label{fig:1}
\end{figure}

To illustrate this point we present the $P_T$ values integrated over
$10^\circ$--intervals as a function of $\theta$ in
Fig.~\ref{fig:1}.
The value of $P_T$ associated with a bin
$\theta_1<\theta<\theta_2$ is determined as
\begin{equation}\label{theta-definition}
  P_T(\theta_1<\theta<\theta_2)
  = \frac{
      \int\limits_{\mbox{\scriptsize cuts, }\theta_1<\theta<\theta_2}
        \rho_T\,dx\,dy
    }{
      \int\limits_{\mbox{\scriptsize cuts, }\theta_1<\theta<\theta_2}
        \rho_0\,dx\,dy
    }\;,
\end{equation} 
where the additional relevant cuts are $E_\gamma>20~\mathrm{MeV}$,
$E_\mu>200~\mathrm{MeV}$.  We evaluate $P_T$ for four different sets
of the form factors corresponding to the boundaries of the allowed
intervals~\eqref{bound-on-sum-from-Kl2gamma},
\eqref{bound-on-dif-from-Kl2gamma}.  One can see that the predictions
for $P_T$ differ significantly, and in the region of large angles,
$\theta>120^\circ$, {\it the sign of $F_V+F_A$ can be unambiguously
determined from the $P_T$ analysis only}.
Suppression of $K_{\pi2}$ decays becomes less efficient 
 at very large $\theta$ angles because of low
efficiency of 
the detection of low energy photons from asymmetrical $\pi^0$ decays, 
so we adopt the region $120^{\circ} < \theta <
160^{\circ}$ as the realistic region with the best sensitivity to the
kaon form factors. The difference between the  $P_T$ values 
increases with $\theta$ and
reaches its maximum value at $\theta\sim150^\circ$, as seen from 
Fig.~\ref{fig:1}.    Although the
differential $K_{\mu2\gamma}$ branching ratio decreases at large
$\theta$, its value is still reasonable at large $\theta$, see
Figure~\ref{fig:2}.  The fraction of the $K_{\mu2\gamma}$ events
within $120^{\circ} < \theta < 160^{\circ}$ is about 7\%.  By making
use of Eq.~(\ref{eq:pt}), the statistical sensitivity to $P_T$ in this
Dalitz plot region is estimated to be
\begin{equation}
 \delta P_T(120^\circ<\theta<160^\circ) \sim
 1.0\times10^{-4}\;,~~~1\sigma~{\rm level}.
\label{statsens}
\end{equation}

To unambiguously determine the sign of $F_V+F_A$ without any
additional data on the difference of the form factors one has to be
able to distinguish the $P_T$ values with $F_V+F_A=0.165$,
$F_A-F_V=-0.24$ and $F_V+F_A=-0.165$, $F_A-F_V=0.04$ (the two closest
lines in Fig.~\ref{fig:1}). For these two sets of the form factors,
the difference between $P_T$ values in the region
$120^\circ<\theta<160^\circ$ is $1.6\times10^{-4}$, that allows
determination of the sign of the sum of the form factors at the
level of 1.6 $\sigma$. Note, there is no usual degeneracy related to
the signs of the sum of the form factors, that provides a
possibility to pin down the sign of the sum without fixing the
difference.

If the difference between the form factors is measured with better
precision than in Eq.~\eqref{bound-on-dif-from-Kl2gamma} one has to
compare the situations with equal $F_A-F_V$ and different signs of the
sum of the form factors, that gives the difference in $P_T$ values in
the same region $(120^\circ<\theta<160^\circ)$ of
$(2.6\div3.6)\times10^{-4}$ depending on the precise value of
$F_A-F_V$. Then the statistical sensitivity~(\ref{statsens}) allows to
determine the signs at the level of $3\div4$ $\sigma$.

So, measurement of $P_T$ allows  
signs of the form factors to be distinguished. It is worth to note in 
passing, that at large $\theta$ the
main contribution to the   $P_T$~(FSI) 
comes from the Dalitz plot region $x>0.4$, where the
uncertainty associated with the momentum of the lepton pair $Q^2$ is
$\lesssim$15\%. This uncertainty is 
small enough to be neglected, because it is significantly less than
the difference in $P_T$ for the form factors of different signs, 
see Fig.~\ref{fig:1}. 

Recently we found~\cite{Bezrukov:2003ka} that the normal muon
polarization $P_N$ is extremely sensitive to the values of the form
factors. Since measurement of $P_T$ pins down the sign of the form
factors, it will provide a possibility of an independent cross-check
of the results from measurement of $P_N$ in the future
experiment~\cite{Kudenko:yk}, which will determine $F_V$ and $F_A$
with accuracy as high as 1\%~\cite{Bezrukov:2003ka} basing only on its
own experimental data.

\section{New Physics}
\label{SecIV}

\subsection{Standard parameterization}
\label{st-param}
For a general investigation of possible contribution 
of some new $CP$-violating sources to the transverse muon polarization in 
$K_{\mu2\gamma}$  we
first introduce the most general four-fermion interaction  
\begin{equation*}
\begin{split}
  {\cal L}&=-\frac{G_F}{\sqrt{2}}V_{us}^*\bar{s}\gamma^{\alpha}
    (1-\gamma_5) u\cdot
    \bar{\nu}\gamma_{\alpha}(1-\gamma_5)\mu
  \\
  &+G_V \bar{s}\gamma^{\alpha}u\cdot
    \bar{\nu}\gamma_{\alpha}(1-\gamma_5)\mu+G_A \bar{s}\gamma^{\alpha}\gamma_5 u\cdot
    \bar{\nu}\gamma_{\alpha}(1-\gamma_5)\mu
  \\
  & +G_S \bar{s}u\cdot \bar{\nu}(1+\gamma_5)\mu 
    +G_P \bar{s}\gamma_5 u\cdot \bar{\nu}(1+\gamma_5)\mu \nonumber 
    +\mathrm{h.c.}\;,  
\end{split}
\end{equation*}
where $G_S$, $G_P$, $G_V$, and $G_A$, arising from new physics, denote 
scalar, pseudoscalar, vector, and axial-vector coupling constants, 
respectively.  Their contribution to $K_{\mu2\gamma}$ decay amplitude
may be taken into account by the redefinition of the usual kaon form
factors as follows,
\begin{eqnarray*}
  f_K &\to& f_K \left(1+\Delta_P +\Delta_A\right),
  \nonumber\\
  F_V &\to& F_V (1+\Delta_V),
  \nonumber\\
  F_A &\to& F_A (1-\Delta_A),
\end{eqnarray*}
with
\begin{equation*}
  \Delta_{(P,A,V)}=
    \frac{\sqrt{2}}{G_FV^*_{us}}
    \left(\frac{G_PB_0}{m_{\mu}},G_A,G_V\right)\,.
\end{equation*}
The constant $B_0$ is related to
quark condensate as $\la 0|\bar{q}q|0\ra=-\frac{1}{2}B_0f_{\pi^0}^2$ and may be
evaluated from the masses of kaon and quarks, 
$B_0=M_{K^0}^2/(m_d+m_s)\approx2$~GeV. 
The scalar type interaction $G_S$ does not contribute to
$K_{\mu2\gamma}$ decay because of parity, and will not be considered
below.  Note, however, that it contributes to the transverse muon
polarization in the $K\to\pi^0\mu\nu_\mu$ ($K_{\mu3}$) decays (while the
pseudoscalar interaction $G_P$ does not).

The imaginary parts of the new coupling constants are responsible for
$CP$-violation and could give rise to the tree-level contributions to 
the muon polarization~\cite{Gabrielli:1992dp,Bijnens:1992en,Chen:1997gf},
\begin{equation*}
\begin{split}
  \rho_T(x,y)=&-2e^2G^2_F |V_{us}|^2 m^2_K m_{\mu}\frac{1-\lambda}{\lambda}
    \sqrt{\lambda y-\lambda^2-r_{\mu}}\\
  & \times\Biggl\{ \Imag[f_K(F_V+F_A)^*]\frac{\lambda}{1-\lambda}
    \times\left(1-x-\frac{r_{\mu}}{\lambda}\right)\\
  & \hphantom{\times\Biggl\{\,}+
    \Imag[f_K(F_V-F_A)^*]\Biggr\}\,.
\end{split}
\end{equation*}
It is convenient to rewrite $P_T(x,y)$ as
\begin{equation*}
  P_T(x,y) = P^V_T(x,y)+P^A_T(x,y)
\end{equation*}
with
\begin{eqnarray}
  P^V_T(x,y)&=&\sigma_V(x,y)\Imag(\Delta_A+\Delta_V)\,,
  \nonumber\\
  P^A_T(x,y)&=&[\sigma_V(x,y)-\sigma_A(x,y)]\Imag(\Delta_P)\,,
  \label{pt-va}
\end{eqnarray}
where
\begin{multline*}
  \sigma_V(x,y)=2e^2G^2_F|V_{us}|^2 m^2_K m_{\mu}f_KF_V
  \\
  \times\frac{\sqrt{\lambda y-\lambda^2-r_{\mu}}}{\rho_0(x,y)}
  \left[\frac{\lambda-1}{\lambda}-
  \l1-x-\frac{r_{\mu}}{\lambda}\r\right]\;,
\end{multline*}
\begin{multline*}
  \sigma_A(x,y)=2e^2G^2_F |V_{us}|^2 m^2_K m_{\mu}f_K F_A
  \\
  \times\frac{\sqrt{\lambda y-\lambda^2-r_{\mu}}}{\rho_0(x,y)}
  \left[\frac{\lambda-1}{\lambda}+
  \left(1-x-\frac{r_{\mu}}{\lambda}\right)\right]\;.
\end{multline*}

\subsection{Distribution over the Dalitz plot}

\begin{figure*}
  \begin{center}
    \includegraphics[width=\textwidth]{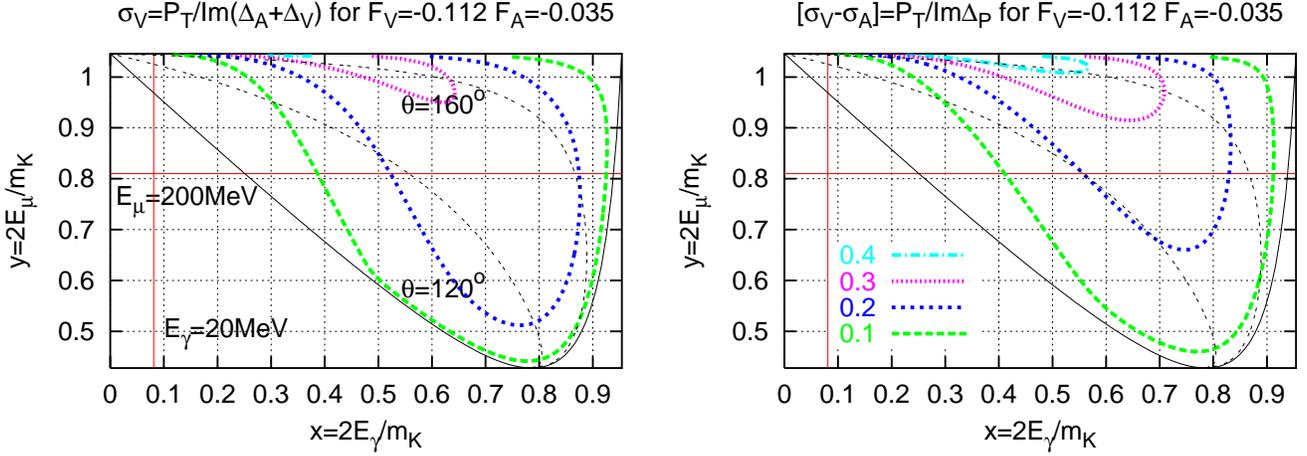}
  \end{center}
  \caption{Distribution of $\sigma_V$ and $[\sigma_V-\sigma_A]$ 
over the Dalitz plot. 
\label{sv_sv-sa}}
\end{figure*}

\begin{figure}
  \begin{center}
    \includegraphics[width=\columnwidth]{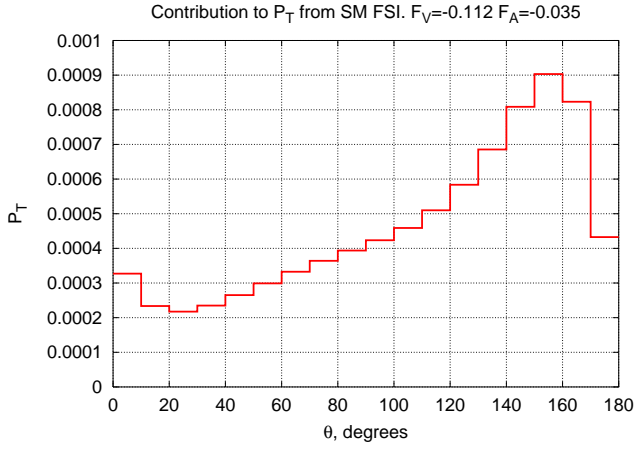}
  \end{center}
  \caption{FSI contribution to $P_T$  for $10^\circ$ wide
    bins in $\theta$; the kaon form factors are equal to the measured
    values~(\ref{measured-values}).}
  \label{fig:FSI-fvfa-exp}
\end{figure}

\begin{figure}
  \begin{center}
    \includegraphics[width=\columnwidth]{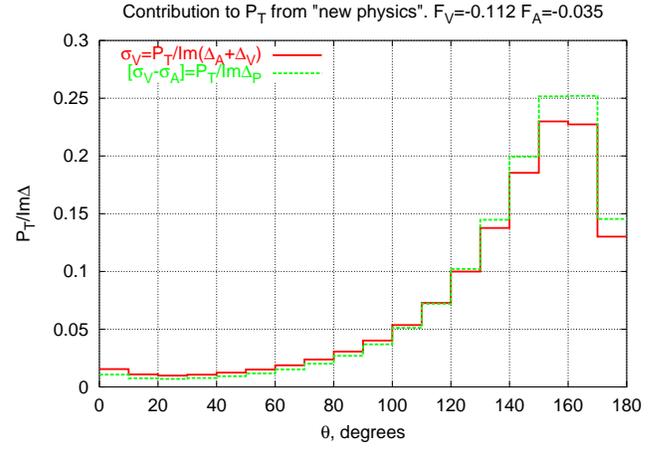}
  \end{center}
  \caption{$P_T$  for $10^\circ$ wide
    bins in $\theta$ at $\Imag(\Delta_V+\Delta_A)=\Imag(\Delta_P)=1$; 
    the kaon form factors are equal to the measured
    values~(\ref{measured-values}).  
  \label{fig:new-phys}}
\end{figure}

To illustrate the sensitivity of the  transverse muon polarization to
new physics we adopt  the experimental
values~(\ref{measured-values}) for the kaon form factors  
and plot the integrated over
$10^\circ$ $\theta$-intervals $P_T$ values  as a function of $\theta$ (see the
definition~(\ref{theta-definition})) in Fig.~\ref{fig:FSI-fvfa-exp}.
The distribution of $P_T^V$ and $P_T^A$ over the Dalitz plot are
completely determined by the functions $\sigma_V(x,y)$ and
$[\sigma_V(x,y)-\sigma_A(x,y)]$, presented in Fig.~\ref{sv_sv-sa}.
Indeed, the values of new $CP$-violating coupling constants provide
only the normalization factors, see Eqs.~(\ref{pt-va}).  To understand
the general behavior we present  $P_T^V$ and
$P_T^A$ values for $10^\circ$ wide bins in $\theta$ at
$\Imag(\Delta_V+\Delta_A)=\Imag(\Delta_P)=1$ in Fig.~\ref{fig:new-phys}.

Comparing Figs.~\ref{fig:FSI-fvfa-exp} and \ref{fig:new-phys} one can
realize that while both FSI and new-physics contributions are peaked
at large $\theta$, the slope is steeper in the former case:
\begin{gather}\label{hierarchy}
  \frac{P_T^V(\theta\sim150^\circ)}{P_T^V(\theta\sim20^\circ)}\simeq20
  \;,\qquad
  \frac{P_T^A(\theta\sim150^\circ)}{P_T^A(\theta\sim20^\circ)}\simeq35
  \;,\\
  \frac{P_T^{FSI}(\theta\sim150^\circ)}{P_T^{FSI}(\theta\sim20^\circ)}\simeq4.5
  \;.
\end{gather} 
This  shows that  the sensitivity to new physics increases the region of 
Dalitz plot with large $\theta$ angles. Indeed, 
the best way to distinguish FSI and new-physics contributions is to
compare the $\theta$-dependence and $P_T$ averaged over the large 
angles, $\theta=120^\circ\div160^\circ$. 
For this range we obtain
\begin{eqnarray}
  P_T^V(120^\circ<\theta<160^\circ)&=&0.14\cdot\Imag(\Delta_V+\Delta_A)
  \;,\label{ptv}\\
  P_T^A(120^\circ<\theta<160^\circ)&=&0.15\cdot\Imag(\Delta_P)
  \;.\label{pta}
\end{eqnarray}
Taking into account the expected experimental sensitivity
\eqref{statsens} one may hope to detect effects of new physics leading to
$\Imag(\Delta_V+\Delta_A)$ or $\Imag\Delta_P$ as small as
$0.7\times10^{-3}$ at  1 $\sigma$ level. 

In the estimation of the sensitivity presented above we neglected the
contribution to $P_T$ from the FSI.  This is possible if one reduces the
uncertainty in the prediction of $P_T^{FSI}$ to values smaller than 
$\delta P_T$.  To this end careful determination of the $F_A$ and
$F_V$ form factors from other experiments is needed. For example, from
$K\to l\nu_le^+e^-$, see Eq.~\eqref{measured-values}, or from the
measurement of the normal muon polarization $P_N$ in $K_{\mu2\gamma}$
decay, which is also a very sensitive to the form factors
observable~\cite{Bezrukov:2003ka}. Simultaneous measurement of both 
$P_N$  and $P_T$ in one experiment with the same statistical sensitivity 
provides  good  opportunity for accurate determination of  $F_A$ and
$F_V$ from $P_N$ and, therefore,  precise calculation of the FSI contribution 
to  measured $P_T$ and clear separation the  non-SM $P_T$ from 
physics background. 

The values of $\Delta_V$, $\Delta_A$ and $\Delta_P$ are completely
determined by the physics beyond the Standard Model.  Below we
consider several extensions of the SM and present the corresponding
bounds on $\Delta_V$, $\Delta_A$ and $\Delta_P$ from existing
experimental data.

\subsection{Two-Higgs-doublet models with suppressed FCNC}

In a general two-Higgs doublet~(2HD) model~\cite{Lee:iz} up- and down-type
quarks, as well as leptons, couple to both two-Higgs doublets. 
The up- and down-type quark mass matrices cannot be diagonalized
simultaneously, that is the reason for flavor changing neutral current
to be induced at tree level.  The relevant terms
read~\cite{Chen:1997gf}
\begin{multline}
  {\cal L}_{\sm{2HD}}=
    \frac{zm_\mu}{4M_H^2}\bar{s}\sum_i[\tilde{\xi}_{i2}^{D*}V_{ui}^{*}
    \l1-\gamma_5\r-\tilde{\xi}_{i1}^{U*}V_{is}^{*}\l1+\gamma_5\r] u
  \\\label{2HD-lagr}
  \times\bar{\nu}\l1+\gamma_5\r\mu 
\end{multline}
where $M_H$ is the mass of charged Higgs boson, $zm_{l_i}$ are
coupling constants of charged Higgs boson and corresponding leptons
and $\tilde{\xi}_{ij}^{U,D}$ are mixing matrices parameterizing up-
and down-type quark interactions with the lightest neutral Higgs
boson.  This interaction results in the muon transverse polarization
\begin{equation}\label{pt2hd}
  P_T=-[\sigma_V-\sigma_A]\frac{B_0z}{2\sqrt{2}G_FM_H^2|V_{us}|}
  \sum_i\Imag[V_{is}^*\tilde{\xi}_{i1}^U+V_{ui}^*\tilde{\xi}_{i2}^{D*}]\;.
\end{equation}

The measurement of the $b\to s\gamma$ gives
$M_H>315$~GeV~\cite{Gambino:2001ew}, perturbativity bound (see,
e.g.,~\cite{Barger:1989fj}) implies $z<2$~GeV$^{-1}$ and, if the only
relevant terms in Eq.~(\ref{pt2hd}) are $\tilde{\xi}^U_{11}$ and
$\tilde{\xi}^D_{22}$, neutral kaon system constrains them to be
$\lesssim10^{-2}\sqrt{M_H/\mathrm{GeV}}$~\cite{Chen:1997gf}.  With
account of Eq.~(\ref{pta}) this favors the limit
\begin{equation*}
\begin{split}
  |P_T(120^\circ<\theta<160^\circ)|& \lesssim 3\times10^{-2}
   \l\frac{315~\rm{GeV}}{M_H}\r^{3/2}\\
  &\times
   \l\frac{z}{2~\rm{GeV}^{-1}}\r
   \l\frac{
   \Imag[\tilde{\xi}_{11}^U+\tilde{\xi}_{22}^{D*}]}{10^{-2}}\r\;.
\end{split}
\end{equation*}
It is more than 300 times larger than the expected experimental
sensitivity to $P_T$~\cite{Kudenko:yk}.  Note, that without special 
fine-tuning between
$\Imag[\tilde{\xi}_{11}^U]$ and $\Imag[\tilde{\xi}_{22}^{D*}]$, the
strongest limit on $P_T$ comes from the search for
transverse muon polarization in $K\to\pi^0\mu\nu$ decay~\cite{Abe:2002vc}, 
since
the interaction~(\ref{2HD-lagr}) generally  provides both $G_P$ and
$G_S$ of the same order~\cite{Kobayashi:1995cy} (see
Sec.~\ref{st-param}). This yields
\begin{equation*}
  |P_T(120^\circ<\theta<160^\circ)|\lesssim3\times10^{-3}\;
\end{equation*}
at 95\% CL. 

On the other hand, depending on the structure of the mixing matrices
$\tilde{\xi}^U$, $\tilde{\xi}^D$, the lower bound on $P_T$ in 2HD
model can be as large as 0.1.

\subsection{Three-Higgs-doublet models}

In these models three different Higgs doublets ($h_u$, $h_d$, $h_l$)
couple to up-, down-type quarks and leptons, respectively.  
The relevant  interaction
between SM fermions and charged Higgs bosons for $K_{\mu2\gamma}$ decay 
reads 
\begin{equation}\label{THDML}
\begin{split}
  {\cal
  L}=&\frac{\sqrt{G_F}}{\sqrt[4]{2}}\sum_{i=1}^2\biggl\{h_i^+
  [\alpha_i\bar{u}VM_d\l1+\gamma_5\r d
  \\
  +&\beta_i\bar{u}M_uV\l1-\gamma_5\r
  d+\gamma_i\bar{\nu}M_l\l1+\gamma_5\r e]\biggr\}+h.c.\;,
\end{split}
\end{equation}
where $V$ is the CKM matrix, $M_d$, $M_u$ and $M_l$ are diagonal
down-type quark, up-type quark and lepton mass matrices, respectively,
and $\alpha_i$, $\beta_{i}$ and $\gamma_i$ ($i=1,2$) are complex
mixing parameters in Higgs sector.  This interaction results
in~\cite{Kobayashi:1995cy}
\begin{eqnarray}
\nonumber
P_T&=&-[\sigma_V-\sigma_A]\l\frac{m_K^2}{m_1^2}-\frac{m_K^2}{m_2^2}\r\\&\times&
\l\Imag{\gamma_1\alpha_1^*}-\frac{m_u}{m_s}\Imag{\gamma_1\beta_1^*}\r
\label{pt3hd}
\end{eqnarray}
where $m_i$ are the masses of charged Higgs bosons.  The current
experimental constraints on the parameters of the theory (see, e.g.,
Ref.~\cite{Grossman:1994jb} for a collection of relevant formulae)
give the same order bound on $P_T$ as   from~\cite{Abe:2002vc} 
for $P_T$  in $K\to\pi^0\mu\nu$
decay (as far as $G_P$ and $G_S$ emerging from
Lagrangian~(\ref{THDML}) are generally of the same order),
$$
  |P_T(120^\circ<\theta<160^\circ)|\lesssim3\times10^{-3}\;,~~~95\%~{\rm CL}\;.
$$
This upper limit is about 30 times larger than the expected
sensitivity of the new experiment.

\subsection{Supersymmetric models with R-parity}

In the minimal supersymmetric extension of the Standard Model (MSSM)
the relevant four-fermion interactions emerge due to W-boson and
charged Higgs boson exchanges with couplings being enhanced by
squark-gluino loops.

Right-handed current interaction is generated by the diagram with
gluino, stop and sbottom particles running in the loop with left-right
mass insertions both for stop and sbottom.  This gives rise to the
effective interaction~\cite{Wu:1996hi}
\begin{eqnarray}
  {\cal L}_I&\!\!\!=\!\!\!&-C\frac{G_F}{\sqrt{2}}\bar{s}\gamma^\alpha\l1+\gamma_5\r u
    \cdot
    \bar{\nu}\gamma_\alpha\l1-\gamma_5\r\mu+\mathrm{h.c.}\;,~~
\label{efsusy}  
\\\nonumber
  C&\!\!\!=\!\!\!&I_0
    \frac{m_tm_b\l A_t-\mu\cot{\beta}\r\l
   A_b-\mu\tan{\beta}\r}{M^4_{\tilde{g}}} \\\nonumber
  &\!\!\!\times\!\!\!&\frac{\alpha_s[M_{\sm{SUSY}}]}{36\pi} V_{31}^{U_{\sm{R}}}V_{32}^{D_{\sm R}*}V_{33}^{SCKM*}\;,
\end{eqnarray}
where $A_t$ and $A_b$ are the soft supersymmetry breaking trilinear
terms for stops and sbottoms, $\mu$ is Higgs superfield mixing
parameter, $\tan{\beta}$ is the ratio of the two Higgs vacuum
expectation values, $M_{\tilde{g}}$ is gluino mass, $V^{U_{\sm{R}}}$
and $V^{D_{\sm R}}$ are the rotations in the generation space between
up-type right-handed squarks and down-type right-handed squarks and
corresponding quark partners, respectively; $V^{SCKM}$ is the super
CKM matrix associated with W-squark-squark couplings, and $I_0$ is
given by the integral
\begin{equation*}
  I_0\!=\!\int_0^1\!dz_1\int_0^{1-z_1}\!\!\!\!\!\!\!\!\!dz_2\frac{24z_1z_2}{\l 
  1+\l\frac{m^2_{\tilde{t}}}{M^2_{\tilde{g}}}-1\r 
  z_1+\l\frac{m^2_{\tilde{b}}}{M^2_{\tilde{g}}}-1\r z_2\r^2}\;.
\end{equation*}
For W-boson exchange, the interaction~(\ref{efsusy}) provides the polarization 
\begin{eqnarray}
  \nonumber
  P_{T_W}&=&-\sigma_VI_0
  \frac{m_tm_b\l A_t-\mu\cot{\beta}\r\l
  A_b-\mu\tan{\beta}\r}{M^4_{\tilde{g}}} \\
  &\times&\frac{\alpha_s[M_{\sm{SUSY}}]}{18\pi}
  \frac{\Imag[{V_{31}^{U_{\sm{R}}}V_{32}^{D_{\sm R}*}V_{33}^{SCKM*}}]}{|V_{us}|}\;.
\label{ptsusywR1}
\end{eqnarray}
Current experimental constraints on the parameters of the
theory~\cite{Hagiwara:pw} leads to
\begin{equation}\label{PTWestimate}
  |P_{T_W}(120^\circ<\theta<160^\circ)|
  \lesssim 0.8\times10^{-3} \;,
\end{equation}
where we set $m_{\tilde{t}}=m_{\tilde{b}}=M_{\tilde{g}}/2$,
$\tan\beta\simeq50$,
$M_{\tilde{g}}\simeq A_b\simeq\mu\simeq A_t\simeq 500 \mathrm{GeV}$,
and $\Imag[{V_{31}^{U_{\sm{R}}}V_{32}^{D_{\sm R}*}V_{33}^{SCKM*}}]\simeq0.5$.

The charged Higgs boson exchange enhanced by gluon-stop-sbottom loops
gives rise to the effective interaction~\cite{Wu:1996hi}
\begin{gather}
  {\cal L}_{II}\!=\!-C_1\frac{G_F}{\sqrt{2}}\bar{s}\l1+\gamma_5\r u
  \cdot
  \bar{\nu}\l1+\gamma_5\r\mu\nonumber\\
-
  C_2\frac{G_F}{\sqrt{2}}\bar{s}\l1-\gamma_5\r u
  \cdot
  \bar{\nu}\l1+\gamma_5\r\mu
  +\mathrm{h.c.}\;,
  \nonumber\\
 \! \!C_1\!\!=\!\!\frac{\alpha_s[M_{\sm{SUSY}}]}{3\pi}I_1\frac{m_tm_\mu\tan{\beta}}{M_{H^+}^2}
  \frac{A_t\cot{\beta}\!+\!\mu}{M_{\tilde{g}}}
  V_{31}^{U_{\sm{R}}}V_{32}^{D_{\sm L}*}V_{33}^{H*}\;,\nonumber\\
  \!\!C_2\!\!=\!\!\frac{\alpha_s[M_{\sm{SUSY}}]}{3\pi}I_1\frac{m_bm_\mu\tan{\beta}}{M_{H^+}^2}
  \frac{A_b\tan{\beta}\!+\!\mu}{M_{\tilde{g}}}
  V_{31}^{U_{\sm{L}}}V_{32}^{D_{\sm R}*}V_{33}^{H*}\;,
  \nonumber\\\nonumber
  I_1\!\!=\!\!\int_0^1dz_1\int_0^{1-z_1}\!\!\!\!\!\!\!\!\!dz_2\frac{2}{ 
  1+\l\frac{m^2_{\tilde{t}}}{M^2_{\tilde{g}}}-1\r 
  z_1+\l\frac{m^2_{\tilde{b}}}{M^2_{\tilde{g}}}-1\r z_2}\;,
\end{gather} 
where $V^H$ is the mixing in the coupling between charged Higgs and
up-type right-handed and down-type left-handed squarks, $V^{U_{\sm
L}}$ and $V^{D_{\sm L}}$ are the rotations in the generation space between
up-type left-handed squarks and down-type left-handed squarks and
corresponding quark partners, respectively.  This results in
\begin{eqnarray}
\nonumber
  P_{T_H}\!\!\!&=&\!\!\!-\!\!\l\sigma_V-\sigma_A\r\frac{\alpha_s[M_{\sm{SUSY}}]}{3\pi}I_1
  \frac{B_0m_t\tan\beta}{M_{H^+}^2}\frac{A_t\cot{\beta}+\mu}{M_{\tilde{g}}}
\\&\times&\frac{\Imag[{V_{31}^{U_{\sm{R}}}V_{32}^{D_{\sm
  L}*}V_{33}^{H*}}]}{|V_{us}|}\nonumber\\\nonumber&
  -&\l\sigma_V-\sigma_A\r\frac{\alpha_s[M_{\sm{SUSY}}]}{3\pi}I_1
  \frac{B_0m_b\tan\beta}{M_{H^+}^2}\frac{A_b\tan{\beta}+\mu}{M_{\tilde{g}}}
\\&\times&\frac{\Imag[{V_{31}^{U_{\sm{L}}}V_{32}^{D_{\sm
  R}*}V_{33}^{H*}}]}{|V_{us}|}\;.
\label{ptsusywR2}
\end{eqnarray}
With the estimate $\Imag[{V_{31}^{U_{\sm{R}}}V_{32}^{D_{\sm
L}*}V_{33}^{H*}}]\simeq0.5$ and
$\Imag[{V_{31}^{U_{\sm{L}}}V_{32}^{D_{\sm R}*}V_{33}^{H*}}]\simeq0.5$
and the same settings as listed below Eq.~(\ref{PTWestimate}) both
terms in Eq.~(\ref{ptsusywR2}) are of the order $10^{-2}$.

So, $P_{T_W}$ and $P_{T_H}$ are 10 and 100 times larger, respectively,  
than the expected
experimental sensitivity.

Note in passing, that without a special cancellation associated with
light superpartners, charge Higgs boson gives large contribution to
$b\to s\gamma$.  This yields the bound
$M_H>315$~GeV~\cite{Gambino:2001ew}, which decreases the upper limit
on $P_{T_H}$ by an order of magnitude.

\subsection{Supersymmetric models without R-parity}

In this section we consider the supersymmetric extensions of the
Standard Model with the violation of the R-parity and the lepton
number.  The relevant superpotential is given by
\begin{equation*}
  {\cal W}_{\sm{RV}}=\half\lambda_{ijk}L_iL_jE_k^c+\lambda'_{ijk}L_iQ_jD_k^c
\end{equation*}
with $L_i$ and $E^c_i$ being the chiral superfields of lepton doublets
and singlets, and $Q_i$, $D_i^c$ denote chiral superfields of quark
doublets and down-type singlets, respectively.  In this model the
contribution to $K_{\mu2\gamma}$ decay arises due to the interaction
\begin{equation*}
  {\cal
  L}_{\sm{RV}}=-\frac{\lambda_{2i2}^*\lambda_{i12}'}{4M^2_{\tilde{e}_{\sm{L_i}}}}
  \bar{s}(1-\gamma_5)u\cdot\bar{\nu}(1+\gamma_5)\mu\;,
\end{equation*}
where $M_{\tilde{e}_{\sm{L_i}}}$ are masses of the left-handed
sleptons. The resulting
transverse muon polarization reads~\cite{Chen:1997gf}
\begin{equation}\label{ptsusy-w/oR}
  P_T=(\sigma_V-\sigma_A)\Imag[\lambda_{2i2}^*\lambda_{i12}']\frac{B_0}{m_\mu|V_{us}|}
  \frac{1}{2\sqrt{2}G_FM^2_{\tilde{e}_{\sm{L_i}}}}\;.
\end{equation}
The strongest relevant experimental limit
 on the parameters of the model comes from the
measurement of $K_L\to\bar{\mu}\mu$ decay rate~\cite{Choudhury:1996ia}, 
and one can obtain
from Eqs.~(\ref{pta}), (\ref{ptsusy-w/oR})
\begin{eqnarray}
&&|P_T(120^\circ<\theta<160^\circ)| \lesssim 1.5\times10^{-5}
\label{ptsusy-w/oR-numbers}\\
&\times&\Biggl(
\l\frac{\Imag[\lambda_{212}^*\lambda_{112}']}{3.8\times10^{-7}}\r
\l\frac{M_{\tilde{\nu}_1}}{M_{\tilde{e}_{\sm{L_1}}}}\r^2\nonumber\\&+&
\l\frac{\Imag[\lambda_{232}^*\lambda_{312}']}{3.8\times10^{-7}}\r
\l\frac{M_{\tilde{\nu}_3}}{M_{\tilde{e}_{\sm{L_3}}}}\r^2
\Biggr)\nonumber
\end{eqnarray}
with $M_{\tilde{\nu}_i}$ being sneutrino masses.  This contribution
(\ref{ptsusy-w/oR-numbers}) is obviously too small to be detected.  Note in
passing that in models with relevant hierarchy in slepton sector or in
models with some cancellation of the sparticle contributions to
$K_L\to\bar{\mu}\mu$ decay, one could expect $P_T$ in $K_{\mu2\gamma}$
at the level of $10^{-3}$.

\subsection{Left-right symmetric models}

In models with left-right gauge symmetries $SU(2)_L\times
SU(2)_R\times U(1)_{B_L}$~\cite{left-right} the four-fermion
interaction which contributes to $K_{\mu2\gamma}$ decay is given by
\begin{equation*}
  {\cal L}_{\sm{LR}}=-\frac{G_F}{\sqrt{2}}\l\frac{g_R}{g_L}\r\xi
  V_{us}^{R*}\bar{s}\gamma_\mu(1+\gamma_5)u\cdot\bar{\nu}\gamma^\mu(1-\gamma_5)\mu
\end{equation*} 
where $\xi$ is left-right complex mixing parameter, $V^R$ is
right-handed CKM matrix and $g_{L,R}$ are coupling constants of
$SU(2)_L$ and $SU(2)_R$, respectively.  This interaction
results~\cite{Chen:1997gf} in
\begin{equation}\label{ptlr}
  P_T=2\sigma_V\frac{g_R}{g_L}\Imag(\xi V_{us}^{R*})\;.
\end{equation}
With the current experimental constraints on the parameters of the
theory~\cite{Barenboim:1996nd} one can obtain from Eqs.~(\ref{ptv}),
(\ref{ptlr})
\begin{multline*}
  |P_T(120^\circ<\theta<160^\circ)| \lesssim 2\times10^{-3}
  \\
  \times \l\frac{g_R/g_L}{1}\r 
  \Imag\biggl[\l\frac{\xi}{0.033}\r\l\frac{V_{us}^{R*}}{|V_{us}|}\r\biggr]\;,
\end{multline*}
that is 20 times larger than the expected statistical
uncertainty of $P_T$~(\ref{statsens}).    

\subsection{Leptoquark models}

There exist two leptoquark models contributing to $P_T$ in
$K_{\mu2\gamma}$ decay.  The quantum numbers of the leptoquarks under
the Standard Model group are
\begin{equation*}
\begin{split}
\phi_1=&\l3,2,\frac{7}{3}\r\;,~~~({\rm Model~I})\;,\\
\phi_2=&\l3,1,-\frac{2}{3}\r\;,~~~({\rm Model~II})\;,
\label{}
\end{split}
\end{equation*}
and the general couplings involving these leptoquarks are given by
\begin{equation*}
\begin{split}
{\cal
L}_{{\sm{LQ}}_I}=&\l\frac{\lambda_1}{2}\bar{Q}\l1+\gamma_5\r e
+\frac{\lambda_1'}{2}\bar{u}\l1-\gamma_5\r L\r\phi_1+\mathrm{h.c.}\;,\\
{\cal
L}_{{\sm{LQ}}_{II}}=&\l\frac{\lambda_2}{2}\bar{Q}\l1+\gamma_5\r L^c
+\frac{\lambda_2'}{2}\bar{u}\l1-\gamma_5\r e^c\r\phi_2+\mathrm{h.c.}\;.
\label{}
\end{split}
\end{equation*} 
The relevant terms read
\begin{equation*}
\begin{split}
{\cal
L}_{{\sm{LQ}}_I}^{K\mu\nu}=&
\frac{\lambda_1^{22}\lambda_1'^{1i*}}{4M_{\phi_1}^2}\bar{s}
\l1+\gamma_5\r\mu\cdot\bar{\nu}_i\l1+\gamma_5\r u+h.c.\;,\\
{\cal
L}_{{\sm{LQ}}_{II}}^{K\mu\nu}=&
\frac{\lambda_2^{2i}\lambda_2'^{12*}}{4M_{\phi_2}^2}\bar{s}
\l1+\gamma_5\r\nu_i^c\cdot\bar{\mu}^c\l1+\gamma_5\r u
+h.c.\;,
\end{split}
\end{equation*} 
where $M_{\phi_{1}}$ are leptoquark masses, and they 
provide~\cite{Chen:1997gf} 
\begin{eqnarray}
\!\!\!\!\!\!\!\!\!\!\!\!\!\!\!\!P_{T_I}\!\!&=&\!\!-\l\sigma_V-\sigma_A\r\Imag[\lambda_1^{22}\lambda_1'^{1i*}]
\frac{B_0}{m_\mu|V_{us}|}\frac{\sqrt{2}}{8G_FM_{\phi_1}^2}\;,
\label{ptlq1}\\
\!\!\!\!\!\!\!\!\!\!\!\!\!\!\!\!P_{T_{II}}\!\!&=&\!\!-\l\sigma_V-\sigma_A\r\Imag[\lambda_2^{2i}\lambda_2'^{12*}]
\frac{B_0}{m_\mu|V_{us}|}\frac{\sqrt{2}}{8G_FM_{\phi_2}^2}\;.
\label{ptlq2}
\end{eqnarray} 
With the current experimental constraints on the parameters of the
models with leptoquarks, the strongest limit on $P_T$ comes from the
measurement of the $P_T$ in
$K\to\pi^0\mu\nu$ decay~\cite{Abe:2002vc} 
(supposing that $G_P$ and $G_S$ constants are
generally of the same order in this theory), that yields
\begin{equation*}
|P_T(120^\circ<\theta<160^\circ)|\lesssim3\times10^{-3}
\end{equation*}
at 95\% CL. This bound is  larger than
 the expected experimental sensitivity by a factor of 30.

\section{Conclusions}
\label{SecV}
 
The analysis presented in this paper concerns the information which
could be drawn from the experiment on the measurement of the
transverse muon polarization $P_T$ in $K\to\mu\nu\gamma$ decay.

It was found that the FSI $P_T$ distribution over the Dalitz plot is
sensitive to the values of the kaon form factors. The best
sensitivity is exhibited in the region of large angles $\theta$
between muon and photon.  The $P_T$ calculated in this region allows
to pin down the signs of the kaon form factors in spite of the
uncertainties associated with the unknown dependence of the form
factors on the momentum of the lepton pair.  To this end the
statistics of $\geq 10^{9}$ $K_{\mu2\gamma}$ events with large
$\theta > 120^{\circ}$ is required.   

It was recently demonstrated~\cite{Bezrukov:2003ka} that the normal
muon polarization $P_N$ is very sensitive to the signs and the values
of the kaon form factors.  Since $P_N$ emerges at tree level, the
statistics expected in the new experiment proposed at JHF allows to
determine for sure the kaon form factors with a few percent accuracy.
The unknown dependence of the form factors on the momentum of the
lepton pair should be determined by fitting the experimental data on
the $P_N$ Dalitz plot distribution.

Finally, it was shown that possible new-physics contributions and the
FSI contribution to $P_T$ show different behavior over the Dalitz
plot. In particular, at large $\theta$ the new-physics contributions
to $P_T$ grow by one and a half orders of magnitude, that suggests to
consider this region as the most interesting to probe the new physics
responsible for $T$-violating effects.  In most cases the contribution
from new physics can exceed the measurable level of $P_T$ by one or
two orders of magnitude. Moreover, extraction of form factor values from 
$P_N$ allows the background contribution from FSI to the non-SM 
T-violating muon  polarization to be determined with small uncertainty.

It is worth to note, that the signs of pion form factors $F^\pi_V$ and
$F^\pi_A$ have not been measured yet~\cite{Hagiwara:pw}.  The
experimental situation there is similar to kaons, though $F^\pi_V$ has
a definite CVC prediction and the pion form factors are almost
constants over the whole Dalitz plot.  The FSI contributions to $P_T$
is of the order of $10^{-4}$ ($10^{-3}$) for $\pi_{\mu2\gamma}$
($\pi_{e2\gamma}$)~\cite{Bezrukov:2002zc}.  From the analysis given
above, we can suggest that the measurement of the transverse lepton
polarization in $\pi_{l2\gamma}$ would allow to pin down the signs of
the pion form factors, if possible contributions to $P_T$ from physics
beyond the Standard Model are negligible.  This issue will be
considered elsewhere.
 
The work is supported in part by RFBR grant 02-02-17398 and by the
program SCOPES of the Swiss National Science Foundation, project
No.~7SUPJ062239.  The work of F.B.\ is supported in part by CRDF grant
RP1-2364-MO-02.  The work of D.G.\ is supported in part by the INTAS
YSF 2001/2-142.


\end{document}